\documentclass[a4paper,11pt]{article}
\usepackage{pos}

\def\be{\begin{equation}}
\def\ee{\end{equation}}

\begin{document}

\title{Universality in High Energy Collisions of small and large systems}
\author{P. Castorina$^{1,2}$, A. Iorio$^{2}$, D. Lanteri$^{1,3}$, H. Satz$^4$ and M. Spousta$^{2}$}
\affiliation{
\mbox{${}^1$ INFN, Sezione di Catania, I-95123 Catania, Italy.}\\
\mbox{${}^2$ Institute of Particle and Nuclear Physics, Faculty of Mathematics and Physics, Charles University}\\
\mbox{V Hole\v{s}ovi\v{c}k\'ach 2, 18000 Prague 8, Czech Republic}\\
\mbox{${}^3$ Dipartimento di Fisica e Astronomia, Università  di Catania, I-95123 Catania, Italy.}\\
\mbox{${}^4$ Fakult\"at f\"ur Physik, Universit\"at Bielefeld, Germany.}
}

\emailAdd{paolo.castorina@ct.infn.it,speaker}

\abstract{Strangeness enhancement and collective flow are considered signatures of the
quark gluon plasma formation. These phenomena have been detected not only in relativistic heavy ion collisions
but also in high energy, high multiplicity events of proton-proton  and proton-nucleus (``small systems'') scatterings. A universal behavior emerges by considering the parton density in the transverse plane as the dynamical quantity to specify the initial condition of the collisions, which in  $e^+e^-$ annihilation at the available energies is too low to expect collective effects.}

\FullConference{%
  40th International Conference on High Energy physics - ICHEP2020\\
  July 28 - August 6, 2020\\
  Prague, Czech Republic (virtual meeting)}


\maketitle

\section{Introduction}

Recent experimental results in
proton-proton ($pp$) and proton-nucleus ($pA$) collisions~\cite{ALICE:2017jyt,Abelev:2013haa,Khachatryan:2016txc,atlas1,cms1,cms2,PHENIX:2018lia,p1,p2} support the conclusion that the
system created in high energy, high multiplicity
collisions with these ``small'' initial settings is essentially the same as that one produced
with ``large'' initial nucleus-nucleus ($AA$) configurations.

The ALICE collaboration reported~\cite{ALICE:2017jyt} an enhanced production
of multi-strange hadrons, previously observed in $PbPb$
collisions~\cite{ABELEV:2013zaa}, in high energy, high multiplicity $pp$ events, shown in fig.~\ref{fig:1}.(left). 

Moreover, the energy loss in $AA$ collisions was
shown to scale in small and large systems~\cite{loss} and another important similarity among $pp$, $pA$, and $AA$
collisions was identified in several measurements of long-range di-hadron azimuthal correlations~\cite{Aad:2012gla,atlas1,cms2,Khachatryan:2016txc},  indicating universality in
flow-like patterns.

In this contribution, based on ref. \cite{cilss}, the universal behavior will be discussed by considering a dynamical variable,  previously introduced to predict the strangeness
enhancement in $pp$~\cite{cs1,cs2,cs3}, which corresponds to the initial entropy 
density of the collisions and takes into account the
transverse size (and its fluctuations) of the initial
configuration in high multiplicity events.

The next sections are devoted to the strangeness enhancement, mean transverse momentum  and elliptic flow. The final section contains our comments,  clarifying why collective effects cannot arise in
$e^+e^-$ annihilation at LEP (or lower) energies.

\section{Universality in strangeness production and in mean transverse momentum}

\begin{figure}
\centering
\includegraphics[width=5cm,height=5cm]{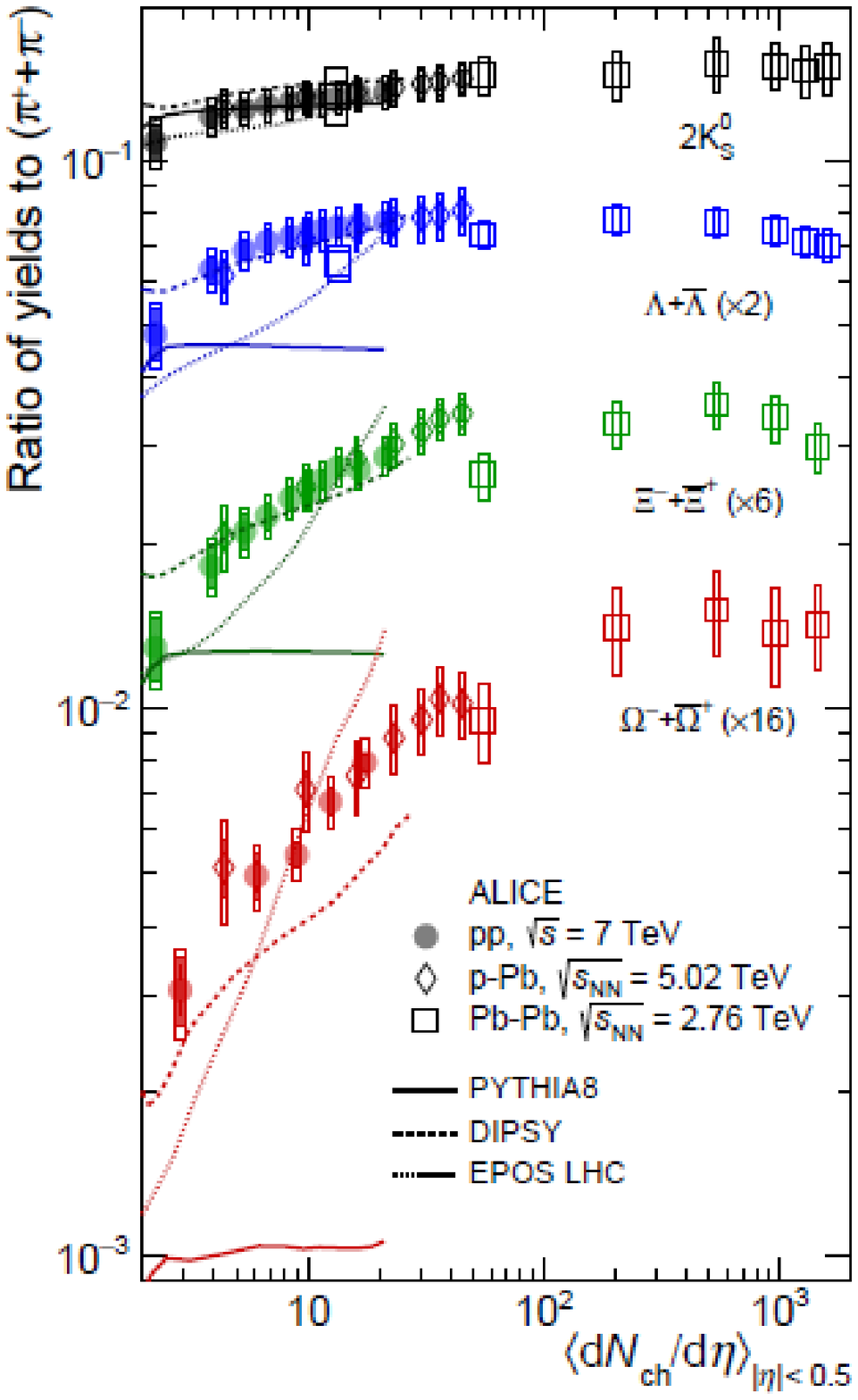}
\quad
\includegraphics[width=4cm,height=5cm]{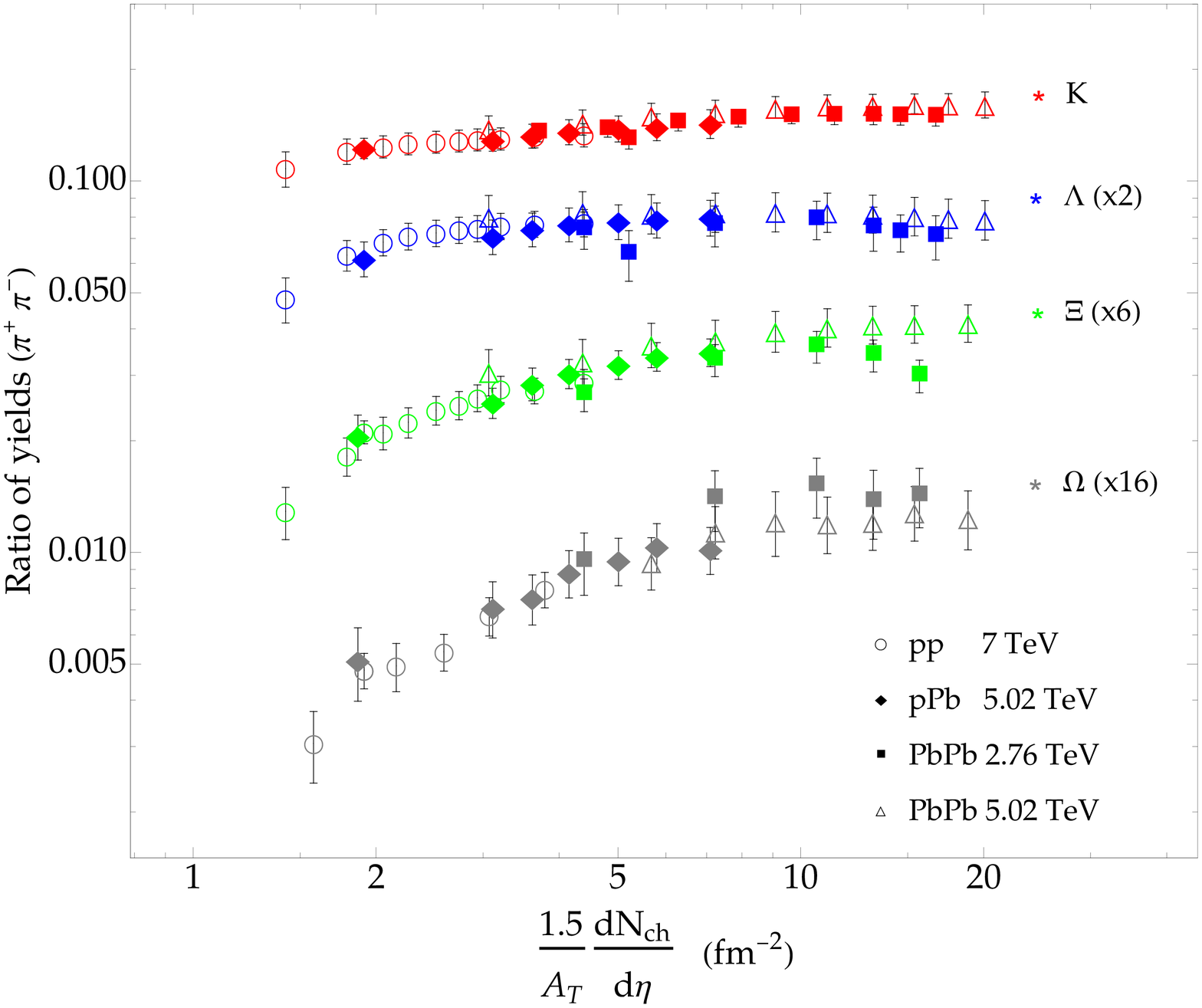}
\caption
{The strangeness production quantified in terms of the ratio of yields of K, $\Lambda$, $\Xi$, and $\Omega$ hadrons to pions evaluated as a function of the multiplicity (left panel) and of the initial 
entropy density (right panel). Data from ~Ref.~\cite{ALICE:2017jyt} (and refs. therein).}
\label{fig:1}
\end{figure}

In fig.\ref{fig:1}.(left) the multi-strange hadron production data are plotted versus the charge multiplicity at mid-rapidity. A stronger evidence of the universal behavior in $pp,pA,AA$ can be discoverd by considering
the initial entropy density $s_0$, given in the one-dimensional hydrodynamic
formulation~\cite{Bj} by the form
\begin{equation}
s_0 \tau_0 \simeq \frac{1.5}{A_T}\; \frac{dN^x_{ch}}{dy} =
\frac{1.5}{A_T}\;\frac{N_{part}^x}{2} \left. \frac{dN^x_{ch}}{dy}\right|_{y=0}\;,
\label{star0}
\end{equation}
with $x \simeq pp$, $pA$, $AA$.
Here $A_T$ is the transverse area, $(dN^x_{ch} / dy)_{y=0}$ denotes the number of produced charged secondaries,
normalized to half the number of participants $N_{part}^x$, in reaction 
$x$, and $\tau_0$ is the formation time.
  The quantity $s_0 \tau_0$ is directly related to the number of partons per unit of transverse area and therefore, 
due to the large fluctuations in high multiplicity events, a reliable evaluation of the transverse area for different collisions as a function of the multiplicity is required.

In studying the strangeness enhancement and the average $p_t$, we use results from Glauber Monte Carlo (MC) ~\cite{Loizides:2017ack} to obtain $A_T$ and the multiplicity for $AA$ and for $pPb$ collisions. For $pp$ scaterings the effective transverse area is sensitive to the fluctuations of the gluon field configurations 
and therefore we apply the CGC parameterization of the transverse size as a function of $N_{ch}$ ~\cite{larry3}. The universal trend in elliptic flow needs a more refined analysis (see sec.~\ref{sec:3}).

Alice data are plotted versus $s_0 \tau_0$ in fig.~\ref{fig:1}.(right), showing a complete smooth trend of the strangeness production among $pp,pA,AA$ data. 

This universal behavior is confirmed by a similar analysis based on the Statistical Hadronization model (SHM) \cite{Becattini1}, where 
strangeness production is
reduced with respect to the predicted rates by one further parameter, $0< \gamma_s \le 1$.
The predicted rate for a hadron species containing $\nu = 1$, $2$, $3$ strange quarks is  suppressed by the factor $\gamma_s^\nu$.

\begin{figure}
\centering
\includegraphics[width=5cm,height=5cm]{CPS-fig1a.eps}
\quad
\includegraphics[width=5cm,height=5cm]{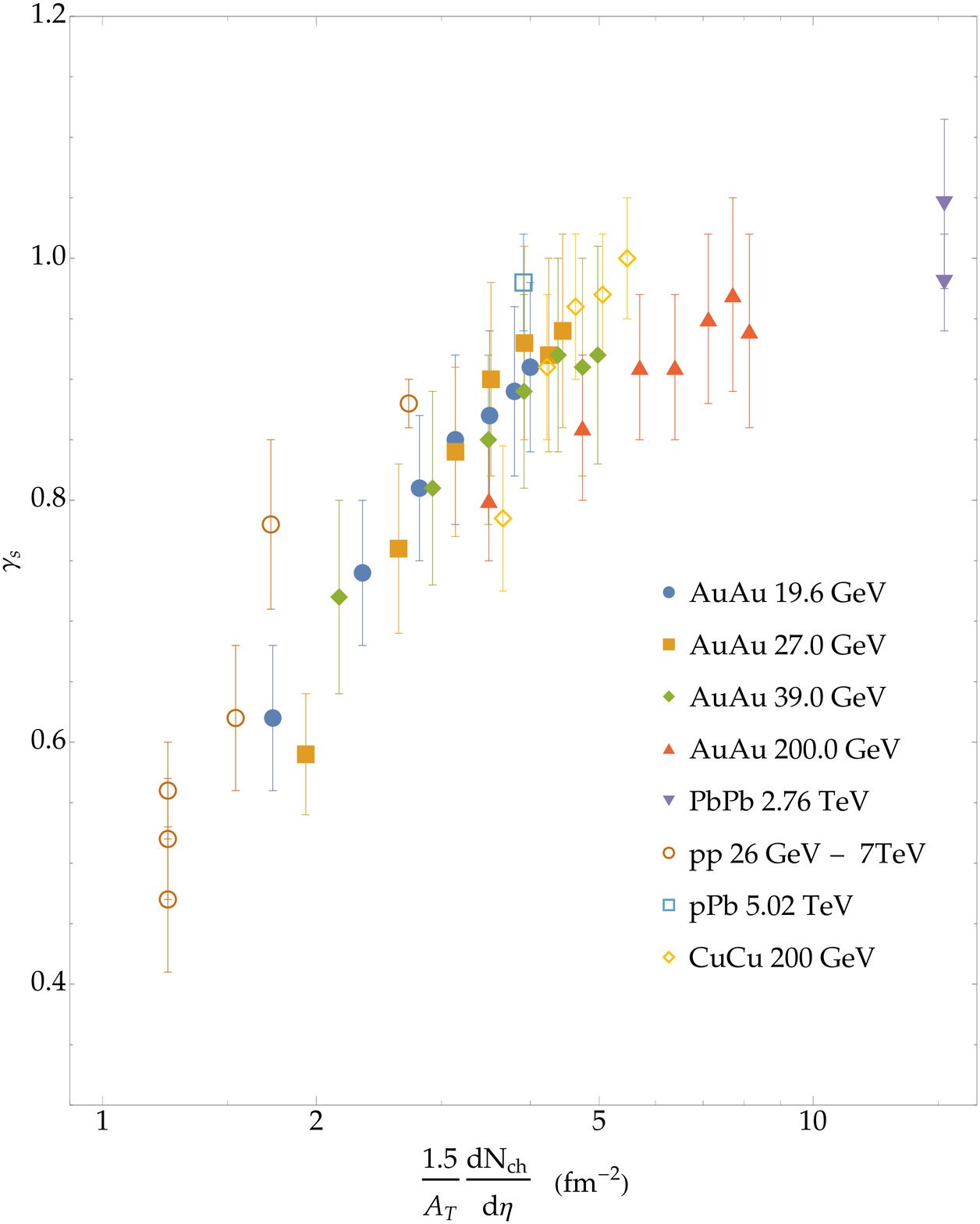}
\caption{The strangeness suppression factor $\gamma_s$ as a function of the energy (left) and of the initial entropy density (right) evaluated for data from Refs.~\cite{Becattini:1996gy,Takahashi:2008yt,Adamczyk:2017iwn}. The Phobos parameterization~\cite{Alver:2010ck} for the relation between charge multiplicity, 
energy and the number of participants is applied for RHIC data.}
\label{fig:2}
\end{figure}

The energy dependence of $\gamma_s$ is reported in fig.~\ref{fig:2}.(left), suggesting a different behavior in strangeness production in $pp$ versus $AA$ collisions.
On the other hand,  by plotting $\gamma_s$ versus the parton density in the transverse plane ( see fig.~\ref{fig:2}.(right)),  a smooth behavior for small and large setting emerges, with a enhancement for $pp$ in high energy, high multiplicity events. The universal trend shows that $\gamma_s$ increases with the parton density in the transverse plane, up to the fixed point $\gamma_s=1$, where any suppression disappears. 

\begin{figure}
\centering
\includegraphics[width=5cm,height=5cm]{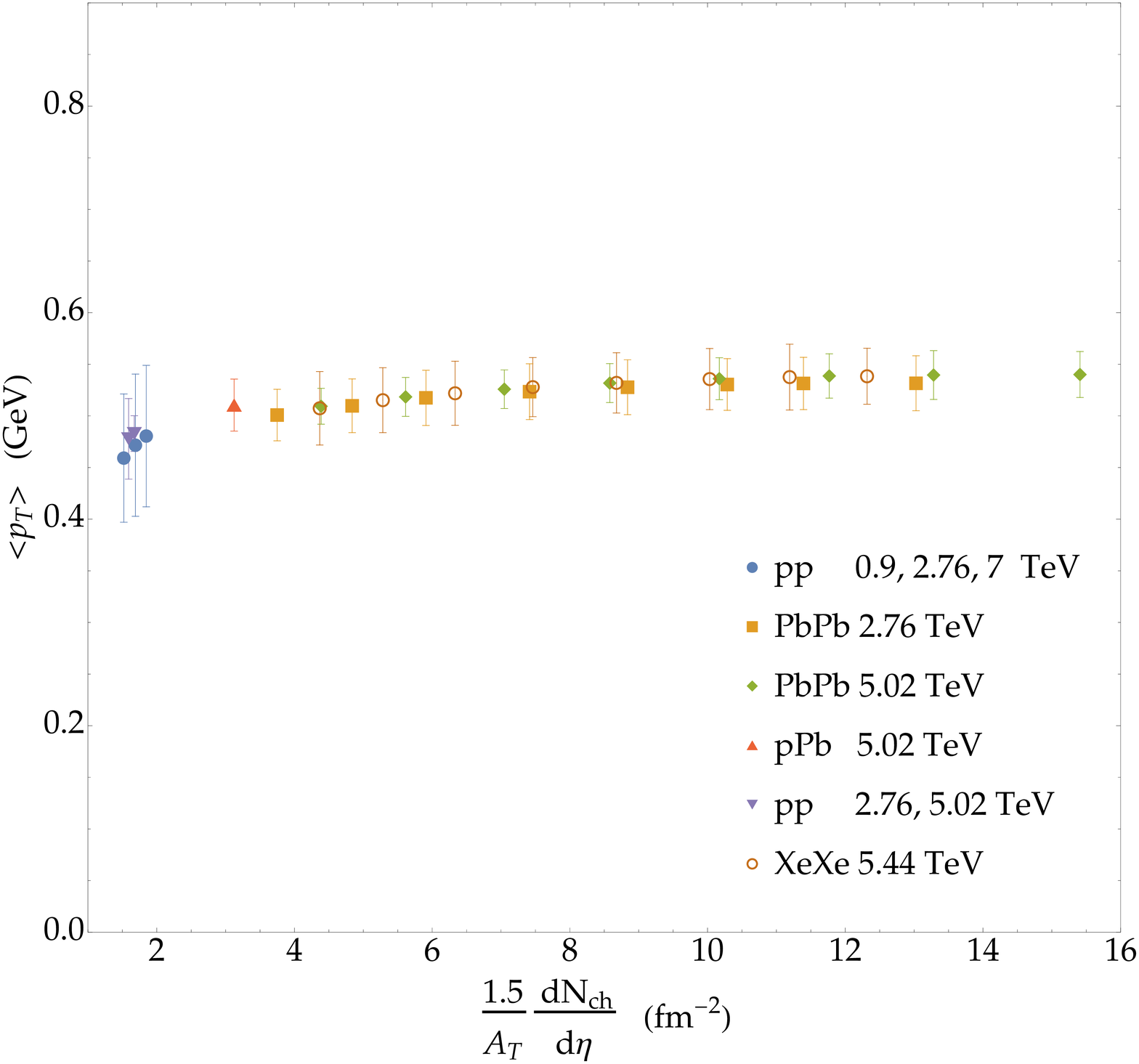}
\quad
\includegraphics[width=5cm,height=5cm]{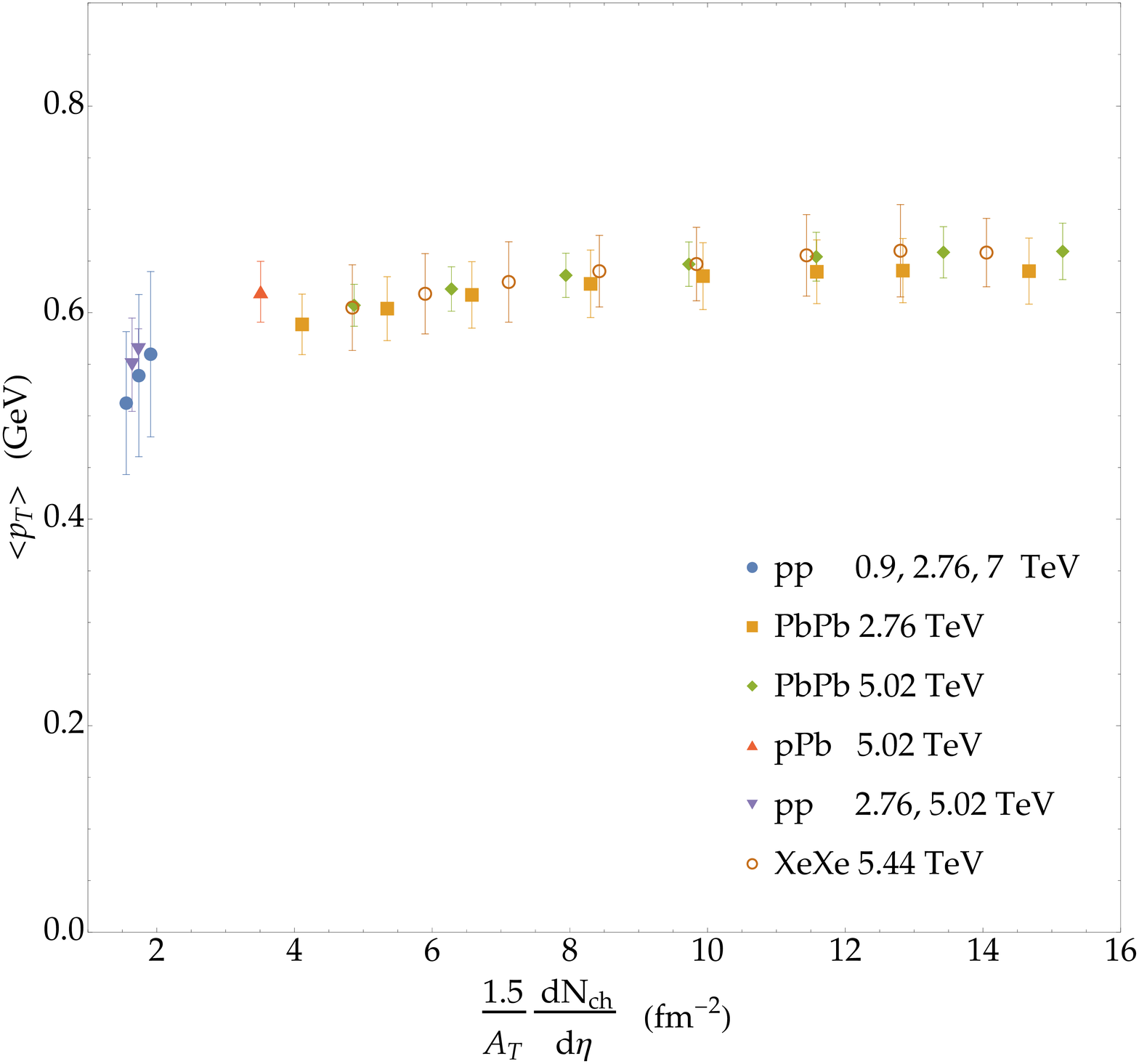}
\caption{Average $p_t$ as a function of initial entropy density evaluated in the interval of $0.15 < p_t < 1.15$~GeV (left) and  $0.15 < p_t < 2$~GeV (\cite{cilss} for refs. to experimental data).} 
\label{fig:3}
\end{figure}

An analogous study can be done for the scaling of the average $p_t$, as discussed in~\cite{larry3}.
We analyze the average $p_t$ in the low transverse momentum region where the soft, non-perturbative, effects in the particle production are more important than in the higher $p_t$ range. 
The behavior of the average $p_t$ is evaluated in the region $0.15 < p_t < 1.15$~GeV (fig.~\ref{fig:3}.(left)) and  $0.15 < p_t < 2$~GeV (fig.~\ref{fig:3}-(right)) for different colliding systems as a function of the previous dynamical variable. One can see that the average $p_t$ for soft particle production follows the same slowly increasing trend for all the collisional systems.

\section{\label{sec:3}Elliptic flow and partecipants eccentricity}
In non-central collisions, the beam direction and the impact parameter vector define a reaction plane for each event. If the nucleon density within the nuclei is continuous, the initial
nuclear overlap region has an ``almond-like'' shape and the impact parameter determines uniquely the initial geometry of the collision. 
In a more realistic description, where the position of the individual nucleons that participate in inelastic interactions is considered, the overlap region has a more 
irregular shape and the event-by-event orientation of the almond fluctuates around the reaction plane. Therefore, in the analysis of 
the elliptic flow where the fluctuations are important, the geometrical eccentricity is replaced by the participant eccentricity, $\epsilon_{part}$, defined using the 
actual distribution of participants. The size of the fluctuation in $\epsilon_{part}$ and its correlated transverse area $S$ (different from the geometrical one, $A_T$) are 
evaluated by Glauber MC.

\begin{figure}
\centering
\includegraphics[width=6cm,height=6cm]{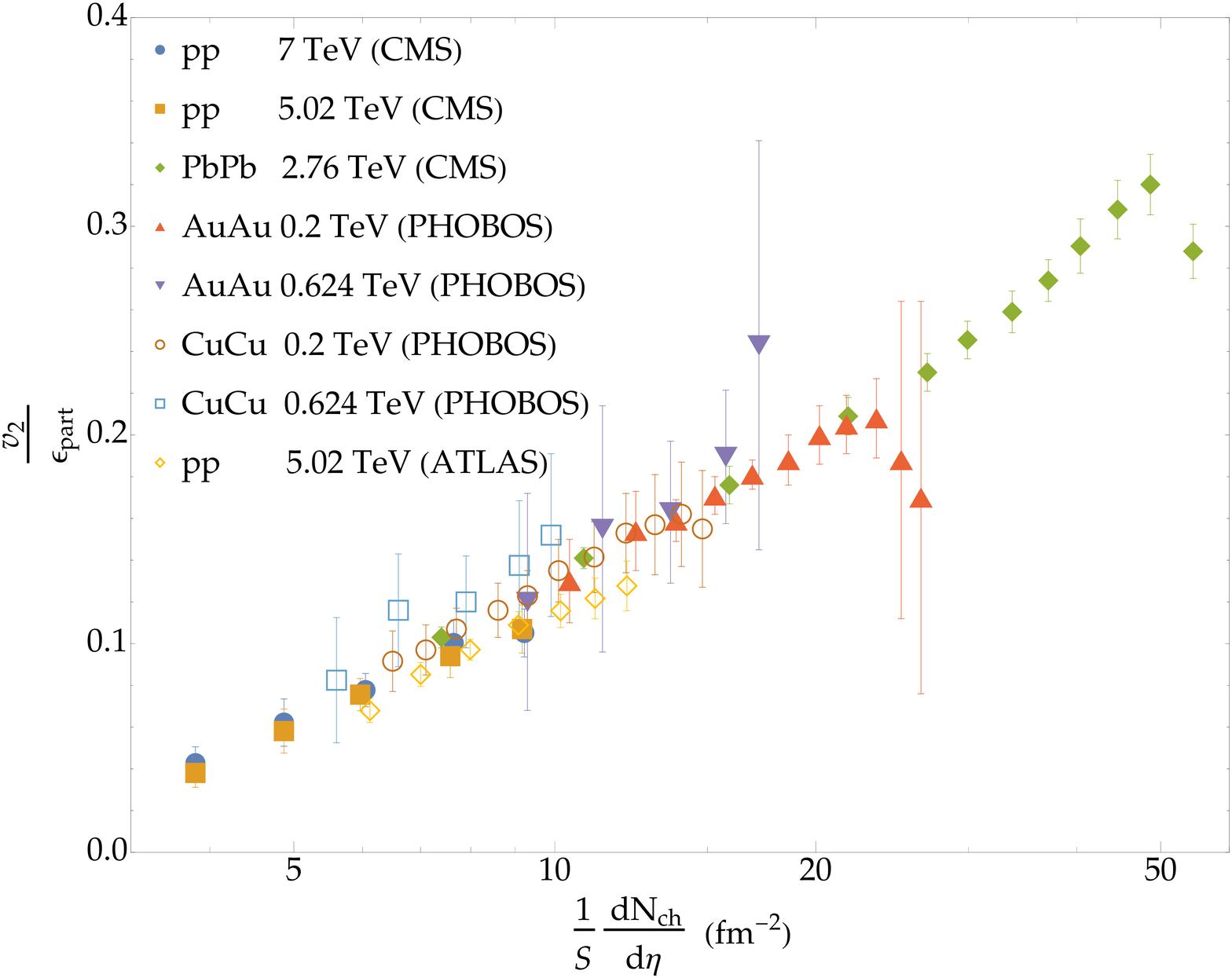}
\quad
\includegraphics[width=6cm,height=6cm]{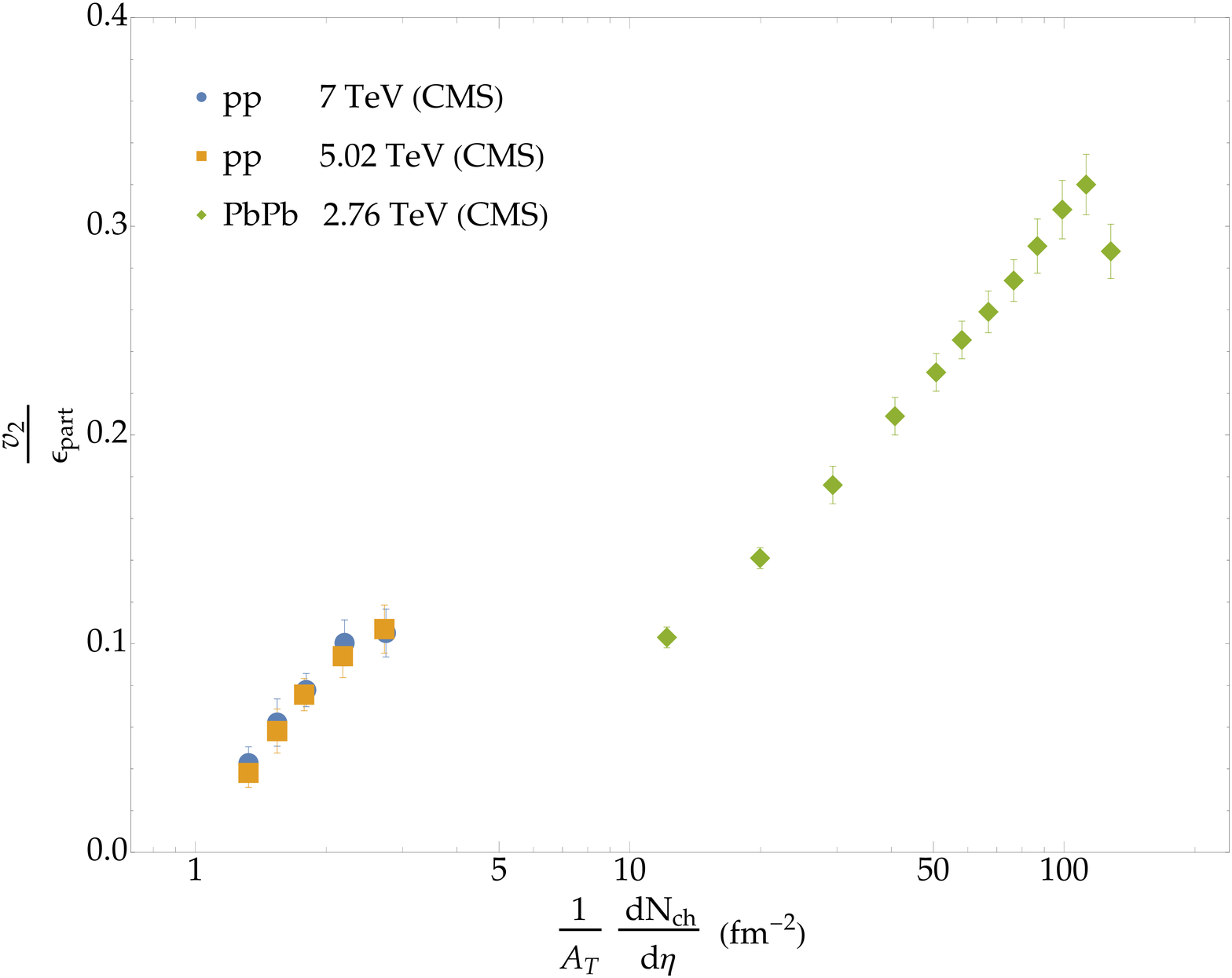}
\caption{The $v_2/\epsilon_{part}$ values for $pp$, $PbPb$, $AuAu$, and $CuCu$ evaluated as a function of entropy density (left) (for data references see \cite{cilss}). The $v_2/\epsilon_{part}$ values for $pp$ and $PbPb$ evaluated as a function of entropy density when the geometrical transverse area  $A_T$, rather tha $S$, is used in the evaluation of the initial energy density for data (right).}
\label{fig:4}
\end{figure}

The scaling of $v_2/\epsilon_{part}$ versus the initial entropy density is depicted in fig.~\ref{fig:4}.(left) for $AA$ and $pp$.
One can see that the $pp$ trend, at lower values, is smoothly followed by the data-points from $AA$ collisions.

To clarify the different role of $A_T$ versus $S$, fig.~\ref{fig:4}.(right)  shows that the scaling in  $v_2/\epsilon_{part}$ is not observed if one considers $A_T$ rather than $S$ in evaluating the initial entropy density.

\section{Comments and Conclusions}
By the scaling variable $s_0 \tau_0$, one can evaluate at which multiplicity the same behavior in 
high-multiplicity $pp$ and $PbPb$ collisions is expected, by solving the equation $(dN/d\eta)_{AA}/A_{T}^{AA} = x/A_T^{pp}(x)$ for $x$ being the multiplicity in $pp$. The result is 
shown in Tab.~2 of Ref.~\cite{cilss}.

A final comment concerns the possible detection of collective effects in $e^+e^-$ annihilation. Indeed, more recently, LEP data have been reconsidered~\cite{badea} to check if a flow-like behavior is generated with this initial, small, non hadronic, setting. The answer is negative and confirmed at lower energy by the BELLE collaboration~\cite{belle}.
Moreover, there is no strangeness enhancement in $e^+ e^-$ annihilation \cite{gammasnoi} as a function of the available energy. 

According to the universality point of view, strangeness enhancement and collective flow are both indications of the formation of an initial system with  high  parton number density in the transverse plane.
Therefore, as confirmed  in Ref.~\cite{clsee},  in $e^+ e^-$ annihilation at LEP or lower energies there is no chance of observing the enhancement of the strangeness production or flow-like effects
because the parton density in the transverse plane is too small.
\vskip10pt
{\bf Acknowledgements} 

P.C. is partially supported by UNCE/SCI/013.

\end{document}